\magnification=\magstep1
\openup 2\jot

\overfullrule=0pt       

\font\cat=cmr7

\def\zed{Z\hskip -3mm Z }
\def\half{\textstyle{1\over2}}
\def\quarter{\textstyle{1\over4}}
 
\input epsf

     
\newcount\refno
\refno=0
\def\nref#1\par{\advance\refno by1\item{[\the\refno]~}#1}
     
\def\book#1[[#2]]{{\it#1\/} (#2).}
     
\def\apj#1 #2 #3.{{\it Ap.\ J.\ \bf#1} #2 (#3).}
\def\cqg#1 #2 #3.{{\it Class.\ Quant.\ Grav.\ \bf#1} #2 (#3).}
\def\jmp#1 #2 #3.{{\it J.\ Math.\ Phys.\ \bf#1} #2 (#3).}
\def\jphysa#1 #2 #3.{{\it J.\ Phys.\ \rm A\bf#1} #2 (#3).}
\def\mpla#1 #2 #3.{{\it Mod.\ Phys.\ Lett.\ \rm A\bf#1} #2 (#3).}
\def\npb#1 #2 #3.{{\it Nucl.\ Phys.\ \rm B\bf#1} #2 (#3).}
\def\phrep#1 #2 #3.{{\it Phys.\ Rep.\ \bf#1} #2 (#3).}
\def\plb#1 #2 #3.{{\it Phys.\ Lett.\ \bf#1\/}B #2 (#3).}
\def\pr#1 #2 #3.{{\it Phys.\ Rev.\ \bf#1} #2 (#3).}
\def\prsa#1 #2 #3.{{\it Proc.\ Roy.\ Soc.\ \rm A\bf#1} #2 (#3).}
\def\prd#1 #2 #3.{{\it Phys.\ Rev.\ \rm D\bf#1} #2 (#3).}
\def\prl#1 #2 #3.{{\it Phys.\ Rev.\ Lett.\ \bf#1} #2 (#3).}
\def\rprog#1 #2 #3.{{\it Rep.\ Prog.\ Phys.\ \bf#1} #2 (#3).}
     
\hbox{ }
\rightline {DTP/96/25}
\rightline {gr-qc/9606002}
\vskip 1truecm
 
\centerline{\bf NON-SINGULAR GLOBAL STRINGS}
\vskip 1truecm
 
\centerline{Ruth Gregory\footnote{$^\spadesuit$}{\sl Email:
dma0rag@gauss.dur.ac.uk}}
\vskip 2mm
\centerline{ \it Centre for Particle Theory, }
\centerline{\it University of Durham, Durham, DH1 3LE, U.K.}
 
\vskip 4mm
\centerline{\cat ABSTRACT}
\vskip 4mm
 
{
\leftskip 10truemm \rightskip 10truemm

\openup -1 \jot

We examine the possibility that time dependence might remove the
singular nature of global string spacetimes. We first show that this
time dependence takes a specific form -- a de-Sitter like expansion
along the length of the string and give an argument for the existence
of such a solution, estimating the rate of expansion. We compare our
solution to the singular Cohen-Kaplan spacetime.
 
\openup 1\jot
}
 
\vskip 1 truecm
{\it PACS numbers: 04.40.-b, 11.10.Lm}
 
{\it Keywords:  gravity, topological defects}
 
\vfill\eject
\footline={\hss\tenrm\folio\hss}
 
\noindent{\bf 1. Introduction.}

Topological defects are ubiquitous in physics, cropping up in some form or
another in such widely disparate fields as string theory and low temperature
physics. Cosmologists in particular have been attracted to 
defects as a possible source for the density perturbations which seeded
galaxy formation[1,2,3]. A topological defect is a discontinuity in the vacuum,
and can be classified according to the topology of the vacuum manifold of
the particular field theory being used to model the physical set up: 
disconnected vacuum manifolds give domain walls, non-simply connected
manifolds, strings, and manifolds with non-trivial 2- and 3-spheres
give monopoles and textures respectively. Strings and monopoles can be
further subdivided into those arising from the breakdown of a local
and global symmetry, being called local and global defects respectively.
However, apart from the (global) domain wall, global defects do not
represent finite energy field configurations, even when we take the
modest definition of `finite' to be finite per unit defect area. 
This indicates that local and global defects will have rather different
behaviour.
Nowhere does this difference show up more dramatically than in the 
coupling of defects to gravity. While local strings[4-6] and 
monopoles[7,8] are
well-behaved, and asymptote flat or locally flat spacetimes, global strings
and monopoles have strong effects at large distances[9,10], and static
global string spacetimes are singular[11,12]. 

In this paper we are interested in the global string. The global
string appears to be in a unique position -- whereas the wall and
global monopole have well-defined non-singular (though not asymptotically
flat) spacetimes, the global string appears to be singular. The
domain wall can appeal to finite energy per unit area, but the mass of 
the global monopole is linearly divergent, therefore it seems strange
that the monopole has a well-defined asymptotic structure whereas the
global string does not. Previous results
on the singularity of the global string metric assumed that the spacetime was
static, not an unreasonable assumption since the local string has a
static spacetime and in order to call the solution a string we might
expect that it has certain features, not least of which a well-defined
and constant physical width. However, as the domain wall shows, staticity
may be too strong an assumption. When the gravitational field of the
domain wall was first explored[4], assuming staticity, it was found to be
singular. Rapidly, this assumption was found to be too restrictive, and
the true, non-static, metric for the domain wall was found to be[13,14]
$$
ds^2 = e^{-4\pi G\sigma |z|} [ dt^2 - dz^2 - e^{4\pi G\sigma t}
(dx^2+dy^2)]
$$
Note the de-Sitter like expansion in the plane of the domain wall.
The experience with domain walls then indicates that perhaps
staticity is too strong an assumption for global strings. 
By including some expansion along the length of the string it may be
possible to reverse the tendency of the global string spacetime to 
collapse in at large distances and allow a non-singular metric. Global
strings would then be acceptable gravitational sources, although
possibly having strong gravitational effects.

A curious corollary of the existence of a non-singular global string 
metric might be its impact on the thermodynamics
of black holes with axion hair[15]. Black holes can carry quantum hair,
and discrete quantum hair[16] was shown to have an effect on the thermodynamics
of a black hole[17]. The mechanism relies on the fact that the theory
admits vortices which interact non-trivially with the `fractional' charge
of the discrete hair by acquiring quantum mechanical phases. Virtual
string worldsheets[18], sitting on the event horizon of the (Euclidean)
black hole then contribute to the Euclidean path integral with different
phases yielding a measurable shift in the Hawking temperature of the black
hole. The problem with detecting the axion hair of Bowick et.al.[15] is
that the appropriate virtual vortices are now global, and give singular
Euclidean geometries[18] and therefore do not appear in the path integral.
If global strings
can be shown to have non-singular spacetimes, then it re-opens the 
possibility that there might be a way of detecting axion hair.

Very little is known about non-static string spacetimes, most research being
directed toward an exploration of the gravity of local strings in an
expanding universe[19,20]. The most detailed study of non-stationary
strings by Shaver and Lake[21] unfortunately assumes that the radial
and azimuthal stresses of the string vanish, which totally precludes its
applicability to global strings, which asymptotically have such stresses
being equal in magnitude to the energy density. 
More recently, Banerjee et.al.[22] considered
the possibility of non-static global strings, however, as we will show, 
they missed a large class of potential solutions, and did not comment
on the singularity structure or otherwise of their metrics.

The purpose of this paper is to address the question of whether a 
non-singular non-static spacetime exists for a global vortex. 
The layout of the paper is as follows. In the next section we
obtain a set of criteria which a physical global string must satisfy
and then derive the general metric and field equations for such a global
string. In section three, we show that the metrics contain a free
parameter, $b_0$, which can be thought of as an effective cosmological
constant along the length of the string, for example, $b_0>0$ gives
de-Sitter like time dependence along the length of the string.
We then exclude regions of parameter space by proving that they lead to 
singular solutions. In section four we give a dynamical systems analysis
of the equations for the global string exterior to the core, and, by
reference to the Cohen-Kaplan solution argue that there does indeed exist
a unique value of $b_0$ for which a non-singular solution exists.
In section five we consider properties of this solution and conclude.

\vskip 5mm

\noindent{\bf 2. Cylindrically symmetric spacetimes.}

\vskip 2mm

In this section we derive the metric and field equations appropriate to
an isolated gravitating U(1) global vortex. We use a ``mostly minus'' 
signature.  Since we are looking for a self-gravitating global string
spacetime, the energy momentum tensor will be taken to be derived
purely from the global string Lagrangian:
$$
{\cal L} = (\nabla_\mu \Phi)^\dagger \nabla ^\mu \Phi - 
{\lambda \over 4} ( \Phi^\dagger\Phi -\eta^2)^2
\eqno (2.1)
$$
By writing
$$
\Phi = \eta X e^{i\chi}
\eqno (2.2)
$$
we reformulate the complex scalar field into two real interacting
scalar field, one of which ($X$) is massive, the other ($\chi$) being the
massless Goldstone boson. 
$$
{\cal L} = \eta^2 (\nabla_\mu X)^2 + \eta^2 X^2 (\nabla_\mu \chi)^2 -
{\lambda \eta^4 \over 4} (X^2-1)^2
\eqno (2.3)
$$
A vortex solution is characterised by the
existence of closed loops in space for which
$$
{1\over 2\pi} \oint {d\chi\over dl} dl = n \in {\rm \zed}
\eqno (2.4)
$$
In other words, the phase of $\Phi$ winds around $\Phi=0$ as a closed loop
is traversed. This in turn implies that $\Phi$ itself has a zero within
that loop, and this is the core of the vortex. 
From now on, we shall assume $n=1$, and look for a solution describing an 
infinitely long isolated straight string.

Our starting point will be that the string spacetime will be expected
to exhibit cylindrical symmetry, namely, it is invariant under rotation about,
and translation along a symmetry axis. Physically, this symmetry axis 
corresponds to the core of the string, and the massless scalar field
corresponds to the azimuthal angle around this symmetry axis. If we
additionally require that the string have fixed proper width, we may
then choose coordinates such that the metric is
$$
ds^2 = e^{2A} dt^2 + 2Fe^{2A} dt dr - dr^2 - e^{2B} dz^2 - C^2 d\theta^2
\eqno (2.5)
$$
and $X=X(r)$. Examining the equation of motion for $X$ indicates that
$F{\dot B}=0$, $C$ and $Fe^A$ are functions of $r$, and $A$ and $B$ 
are separable. The form of the energy-momentum
tensor and Einstein tensor then indicates that we must take $F=0$. Since
$A$ is separable, we can always redefine $t$ to absorb any time
dependence of $A$, thus obtaining
$$
ds^2 = e^{2A(r)} dt^2 - dr^2 - e^{2B(r,t)} dz^2 - C^2(r) d\theta^2
\eqno (2.6)
$$
as the final form for the general non-static metric for the global
string.  
The Einstein tensor for this metric is
$$
\eqalignno{
G^r_r &= (\ddot{B} + {\dot B}^2 ) e^{-2A} - A'B' - (A'+B') {C'\over C}
& (2.7a) \cr
G^0_0 &= - \left [ B'' + {B'}^2 + {C''\over C} + {B'C'\over C} \right ]
& (2.7b) \cr
G^z_z &= - \left [ A'' + {A'}^2 + {C''\over C} + {A'C'\over C} \right ]
& (2.7c) \cr
G^\theta_\theta &= (\ddot{B} + {\dot B}^2 ) e^{-2A} - \left [ A'' +
 {A'}^2 + B'' + {B'}^2 +A'B' \right ]
& (2.7d) \cr
G^t_r &= {\dot B} A' - {\dot B}' - {\dot B} B' & (2.7e) \cr
}
$$

For the global string fields, recall that $\chi=\theta$, and 
since we are dealing with an isolated global string, we may,
without loss of generality, set $\sqrt{\lambda} \eta =1$ thereby
choosing units in which the string width is order unity.
We will then absorb the symmetry breaking scale into
the parameter $\epsilon = 8 \pi G \eta^2$ which represents the gravitational
strength of the string, generally assumed to be small. 
The equation of motion for $X$ is then:
$$
-X'' - \left [ {C'\over C} + A' + B' \right ] X' + {X\over C^2}
+ {\half} X (X^2-1) =0
\eqno (2.8)
$$
and the energy-momentum tensor of the vortex
$$
\eqalign{
8 \pi G T^0_0 = 8 \pi G T^z_z &= \epsilon \left [ {X'}^2 + {X^2\over C^2} + 
{\quarter} (X^2-1)^2 \right ]  = \epsilon {\hat T}^0_0 \cr
8 \pi G T^r_r &= \epsilon \left [-{X'}^2 + {X^2\over C^2} + 
{\quarter} (X^2-1)^2 \right ]  = \epsilon {\hat T}^r_r \cr
8 \pi G T^\theta_\theta &= \epsilon \left [ {X'}^2 -{X^2\over C^2} + 
{\quarter} (X^2-1)^2 \right ]  = \epsilon {\hat T}^\theta_\theta \cr
}
\eqno (2.9)
$$

In flat space ($\epsilon=0$), the equation of motion reduces to
$$
-X'' - {X'\over r} + {X\over r^2} + {\half}  X (X^2-1) =0
\eqno (2.10)
$$
which does not have a closed analytic solution, but can be integrated
numerically. In particular, note that  as $r\to\infty$, 
the asymptotic form for $X$ is given by
$X\sim 1 - 1/r^2$, and hence $T^0_0 \propto 1/r^2$
giving rise to a logarithmically divergent energy per unit length for
the vortex. 

Now let us consider coupling in gravity ($\epsilon \neq 0$).
Referring to the Einstein equations, it can be seen that boost symmetry
of the energy momentum tensor ($T^0_0 = T^z_z$) implies that 
$A'=B'$. Finally, referring to (2.7e), we see that ${\dot B}'=0$, 
consistent with the separability of $B$:
$$
B(r,t) = A(r) + b(t)
\eqno (2.11)
$$

Therefore, collecting all this information, we see that the general form
of the global string metric is
$$
ds^2 = e^{2A} (dt^2 - e^{2b(t)}dz^2) - dr^2 - C^2(r) d\theta^2
\eqno (2.12)
$$
with field equations
$$
\eqalignno{
A'' + {C''\over C} + {A'}^2 + {A'C'\over C} &= - \epsilon {\hat T}^0_0 
& (2.13a) \cr
({\ddot b} + {\dot b}^2 ) e^{-2A} - {A'}^2 - 2{A'C'\over C}
&= \epsilon {\hat T}^r_r & (2.13b) \cr
({\ddot b} + {\dot b}^2 ) e^{-2A} - 2A'' - 3 {A'}^2 &= \epsilon
{\hat T} ^\theta_\theta & (2.13c)
}
$$
where the ${\hat T}^a_b$ are defined in (2.9),
together with equation (2.8) for $X$.

\vskip 5mm

\noindent{\bf 3. The equations of motion.}

\vskip 2mm

Now we will analyse the equations derived in the previous section
for the isolated global string spacetime and find under what conditions
the solutions are singular.
Clearly (2.13b,c) imply
$$
{\ddot b} + {\dot b}^2 = b_0, \;\; {\rm a\ constant}
\eqno (3.1)
$$
and thus
$$
b(t) = \cases{ \ln \cosh \sqrt{b_0} t\ ;\pm \sqrt{b_0}t & $b_0 > 0 $ \cr
               b_1 \ln t & $b_0 = 0$, $b_1$ a constant \cr
               \ln \cos \sqrt{|b_0|} t & $b_0 < 0 $ \cr}
\eqno (3.2)
$$
Note that the non-static solutions of Banerjee et.al.[22] correspond
to $b_0=0$, $b_1\neq0$. Moreover, since the Einstein equations for
$b_0=0$ are identical in structure whether or not $b_1$ is zero,
we must expect that solutions with $b_1 \neq 0$ are singular since
static global strings are singular, and indeed, Banerjee et.al.~pointed
out that their solutions were a coordinate transformation of the
static solution. Therefore, let us take $b_0$ to be arbitrary, and
consider two cases in turn, $b_0 \leq 0$, and $b_0>0$. In either case,
the system of equations we are dealing with is:
$$
\eqalignno{
[C' e^{2A}]' &= -\epsilon C e^{2A} \left [ {2X^2 \over C^2}
+ {\quarter} (X^2-1)^2 \right ] & (3.3a) \cr
[A'Ce^{2A}]' &= C \left [ b_0 - {\quarter} \epsilon e^{2A} (X^2-1)^2
\right ] & (3.3b) \cr
{A'}^2 + {2A'C'\over C} &= b_0 e^{-2A} - \epsilon \left [
-{X'}^2 + {X^2\over C^2} + {\quarter} (X^2-1)^2 \right ] & (3.3c) \cr
[Ce^{2A} X']' &= {Xe^{2A} \over C} + {\half} C e^{2A} X(X^2-1)
& (3.3d) \cr
}
$$

\vskip 2mm

\noindent {\it (i) $b_0 \leq 0$.}

\vskip 1mm

Note first that (3.3a,b) together imply that
$$
[Ce^{2A}]'' = 2 b_0 C - \epsilon Ce^{2A} \left [ {2X^2 \over C^2}
+ {3\over 4} (X^2-1)^2 \right ] \leq 0
\eqno (3.4)
$$
Thus $Ce^{2A} \leq r$ for all $r$. Next we observe that if $(Ce^{2A})' =0$
at any point, then (3.4) implies that $Ce^{2A}\to 0$ with
$(Ce^{2A})'$ strictly negative at some finite $r$, $r_0$ say. Thus
$2A' + C'/C \to -\infty$ as $r \to r_0$. Since (3.3b)
implies that $A'$ is strictly negative away from $r=0$, we may conclude 
that either $A'C'/C$ or $A'$ (or both) become infinite at $r_0$.
Therefore
$$
R_{abcd}^2 \propto \left ( {C'' \over C} \right ) ^2 +
2 \left ( {A'C'\over C} \right ) ^2 + 2 ( A'' + {A'}^2 )^2
+ ({A'}^2 - b_0 e^{-2A})^2
\eqno (3.5)
$$
would become infinite at $r_0$ indicating a physical singularity.
For a non-singular spacetime, we therefore take $(Ce^{2A})'>0$.
This in turn implies that
$$
1 + \int _0 ^r \left \{ 2b_0 C - \epsilon Ce^{2A} \left [
{2X^2 \over C^2} + {3\over 4} (X^2-1)^2 \right ] \right \}dr
>0 \;\;\; \forall \ r
\eqno (3.6)
$$
Therefore the integrals
$$
\eqalignno{
\alpha_1 &= \epsilon \int_0^\infty { e^{2A} X^2 \over C} dr & (3.7a) \cr
\alpha_2 &= \epsilon \int_0^\infty {\quarter} Ce^{2A} (X^2-1)^2 & (3.7b) \cr
\alpha_3 &= |b_0| \int_0^\infty C & (3.7c) \cr
}
$$
are convergent, and satisfy the inequality
$$
2\alpha_1 + 3 \alpha_2 + 2 \alpha_3 < 1
\eqno (3.8)
$$
In addition, an examination of (3.3d), together with finiteness of (3.7a-c)
shows that
$$
\alpha_4  = \epsilon \int_0^\infty Ce^{2A} {X'}^2 dr 
\eqno (3.9) 
$$

We may now integrate (3.3a,b) out to infinity to obtain
$$
\eqalign{
A'C e^{2A} &\to - (\alpha_2 + \alpha_3) \cr
C' e^{2A} &\to 1 - 2 \alpha_1 - \alpha_2 \cr
}
\eqno (3.10)
$$
as $r\to \infty$, and multiplying (3.3c) by $(Ce^{2A})^2$, we also have
$$
(\alpha_2 + \alpha_3)^2 - 2(\alpha_2 + \alpha_3)
( 1 - 2 \alpha_1 - \alpha_2) = Ce^{2A} \left [
b_0C + \epsilon Ce^{2A} \left ( {X'}^2 - {X^2\over C^2} - {\quarter}
(X^2-1)^2 \right ) \right ] 
\eqno (3.11)
$$
But, finiteness of the integrals in (3.7,9) requires that each of the integrands
separately is $o(r^{-1})$ as $r\to\infty$. Hence $(Ce^{2A})($integrand$) \to
0$ as $r\to\infty$. Thus the r.h.s.~of (3.11) vanishes at infinity
and we have
$$
\eqalign{
(\alpha_2 + \alpha_3) \left [ \alpha_3 + 3\alpha_2 + 4
\alpha_1 - 2 \right ] &=0 \cr
\Rightarrow 2 + 3 (\alpha_2 + \alpha_3) = 2 [ 2\alpha_3 + 3\alpha_2
+2  \alpha_1  ] &<2 \cr
}
\eqno (3.12)
$$
which cannot be satisfied since all of the $\alpha_i$ are positive.

Therefore the spacetime must be singular for $b_0 \leq 0$. Note that
this argument makes no use of the explicit form of $X$, only the general
form of the energy-momentum tensor.

\vskip 2mm
 
\noindent {\it (ii) $b_0 > 0$, preliminary considerations.}
 
\vskip 1mm

Note that, irrespective of the sign of $b_0$, (3.3a) implies that
$[C'e^{2A}]' \leq 0$ and hence that $ C'e^{2A} \leq 1$. Therefore
$C'e^{2A}$ either remains positive, or it does not. 

If $C'e^{2A} >0$ for all $r$, then $2 \alpha_1  + 
\alpha_2 \leq 1 $, where the $\alpha_i$ were defined in (3.7).
Thus, using the asymptotic properties of the integrands appearing in
the $\alpha_i$ as before, we may conclude from (3.3b,c) that
$$
\eqalignno{
[A'C e^{2A}]' &\sim b_0 C & (3.13a) \cr
{A'}^2 &\sim b_0 e^{-2A} & (3.13b) \cr
}
$$
which is solved by
$$
C = C_0 \;\;\; ; \;\;\;\; e^A = \sqrt{b_0}r + A_0
\eqno (3.13c)
$$
where $C_0$ and $A_0$ are constants. But an examination of the explicit
form of the integrals (3.7b,c) shows that this 
solution is inconsistent with the finiteness of $\alpha_1$ and $\alpha_2$.
Therefore $C'e^{2A}=0$ for some finite $r$, $r_0$, say.

Suppose first that $C'=0$, $e^{2A} \neq0$. Then (3.3a) implies $C''<0$,
hence $C'$ becomes negative, and must therefore remain negative.
Since an isolated global string has $X\simeq1$ outside the core, if $C\to 0$
the spacetime will be singular, hence $C$ must be bounded away from zero
for all $r$, in which case $C', C''\to0$ as $r\to\infty$.
(3.3a) (and its first integral) then implies that $A', e^{2A}\to\infty$
which contradicts (3.3c) and in any case would give a singular spacetime.

The only remaining possibility is therefore that $e^{2A}\to0$ at some
finite $r$, $r_0$ say. Non-singularity of the spacetime requires 
additionally that $C'(r_0)=0$. $r_0$ would then represent an event 
horizon for the global string spacetime, analogous to that of the domain
wall spacetime[13,14]. From (3.3a-c), we see that the asymptotic solution
near this point would be
$$
\eqalign{
e^A &\sim \sqrt{b_0} (r_0-r) \cr
C &= C_0 + O(r_0-r)^2 \cr
X &\simeq 1 - {1\over C^2} \cr
}
\eqno (3.14)
$$
In other words
$$
ds^2 \simeq b_0 (r_0-r)^2 [ dt^2 - \cosh^2 \sqrt{b_0}t dz^2]
-dr^2 - C_0^2 d\theta^2
\eqno (3.15)
$$
near $r=r_0$. The curvature invariants for this metric are all finite
and $r=r_0$ does indeed appear to be an event horizon, but before
investigating whether (3.14) is indeed possible as an asymptotic
solution for the global string, we should verify the coordinate nature
of the singularity at $r=r_0$. 

Note that the metric (3.15) is reminiscent of that of Harari and
Polychronakos[23], who considered static cylindrically symmetric
solutions with event horizons. Their metric however had $g_{zz}\equiv1$
and hence was not appropriate to a boost symmetric source. Nonetheless,
we can define similar Kruskal-like coordinates in the vicinity of 
$r=r_0$
$$
\eqalign{
X &= (r_0-r) \cosh \sqrt{b_0}t \cos \sqrt{b_0}z \cr
Y &= (r_0-r) \cosh \sqrt{b_0}t \sin \sqrt{b_0}z \cr
T &= (r_0-r) \sinh \sqrt{b_0}t \cr
}
\eqno (3.16)
$$
in terms of which (3.15) becomes
$$
ds^2 \simeq dT^2 - dX^2 - dY^2 - C_0^2 d\theta^2
\eqno (3.17)
$$
The putative event horizon, $r=r_0$, corresponds to $X^2 + Y^2 = T^2$.
Although the coordinate transformation (3.16) is not one to one,
it can be made so by restricting the range of $z$, to $(-{\pi\over\sqrt{b_0}},
{\pi\over\sqrt{b_0}})$ for example. Therefore, it provides a coordinate system
which extends beyond the event horizon $r=r_0$, and verifies the 
coordinate nature of that singularity. This situation is more
complicated than that of Harari and Polychronakos where $g_{zz}\equiv1$,
and the ``Kruskal'' diagram is shown in figure 1.
\midinsert \hskip 4truecm \epsfxsize=8truecm
\epsfbox{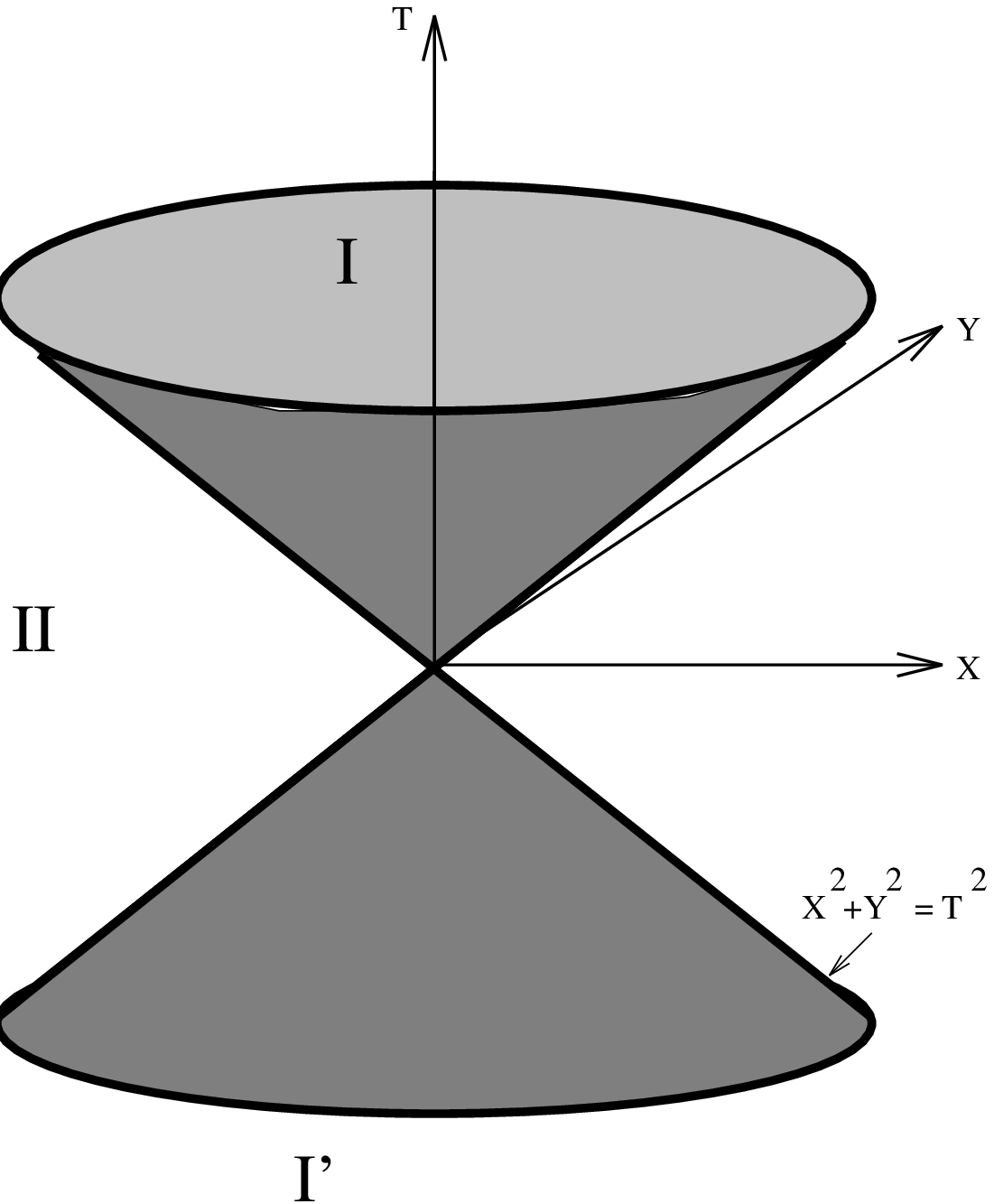} \medskip \hskip 2 truecm
\vbox{ \hsize=11.5 truecm 
\noindent {\cat FIGURE (1): The ``Kruskal'' diagram of the global
string spacetime in $\{X,Y,T\}$ coordinates. Region II, exterior to 
the cone, corresponds to the global string spacetime interior to the
event horizon. Regions I and I' correspond to the spacetime exterior
to the event horizon reached by future and past pointing null geodesics
respectively. }}
\endinsert

To sum up, for the spacetime to be nonsingular, we have shown that $b_0>0$,
and there must exist a $(b_0,r_0)$ such that the solution to (3.3)
asymptotes (3.14). We will address this question in more detail in the 
next section, for the moment, we simply note that (3.3a,b) imply
$$
\eqalign{
2\epsilon \int_0^{r_0} {X^2 e^{2A} \over C} dr + {\epsilon \over4}
\int_0^{r_0} Ce^{2A} (X^2-1)^2 dr &=1 \cr
b_0 \int_0^{r_0} C &= {\epsilon \over4}\int_0^{r_0} Ce^{2A} (X^2-1)^2 dr\cr
}
\eqno (3.18)
$$
Therefore, for $\epsilon \ll 1$, we may approximate $X,C$ and $A$ by their
flat space values, at least out to $r=O(\epsilon^{-{\half}})$, and thus obtain
$$
b_0 < O(\epsilon^2)
\eqno (3.19)
$$

\vskip 5mm

\noindent{\bf 4. The exterior spacetime.}

\vskip 2mm

We would now like to examine whether (3.14) is indeed admissible as an
asymptotic solution of (3.3). In order to address this question, we first
recall the analysis of Cohen and Kaplan[9]. Cohen and Kaplan were 
interested in the form of the global string metric inside its singularity
radius. To get the Cohen-Kaplan (CK) metric, we set $b_0=0$ and take
$X=1$ outside the core. Since we expect $1-X \sim 1/C^2$, for 
$C>\epsilon^{-{1\over2}}$ we expect this to be an excellent approximation,
and for $C>$ few, it should give a good working approximation. Under these
assumptions, (3.3a,b) become
$$
\eqalignno{
[C'e^{2A}]' &= -{2\epsilon e^{2A}\over C} & (4.1a) \cr
[A'Ce^{2A}]' &=0 & (4.1b) \cr
}
$$
Thus 
$$
(e^{2A})' = -2 K \epsilon /C
\eqno (4.2)
$$
where
$$
K = \int {\quarter} C e^{2A} (X^2-1)^2 dr = O(1)
\eqno (4.3)
$$
Rewriting $Cdu=-dr$, one obtains the CK solution:
$$
\eqalign{
e^{2A} &= {u\over u_0} \cr
C^2 &= \gamma \sqrt{{u_0\over u}} \exp \left \{ {u_0^2-u^2 \over u_0}
\right \} \cr
}
\eqno (4.4)
$$
Where $u_0={1\over2K\epsilon}$ is given by (4.2), and $\gamma$ is of order unity
(for convenience, we will take $\gamma=1$).
$u=u_0$ corresponds roughly to the core of the string, and $u=0$ the
singularity, which, can be seen to occur approximately at
$$
r = \int_0^{u_0} Cdu \sim {1\over\epsilon}e^{1\over2\epsilon}
\eqno (4.5)
$$
a very large radius!

Note that the simplicity of the Cohen-Kaplan analysis relied on the vanishing
of the right hand side of (4.1b); if $b_0\neq0$, this r.h.s.\ is equal
to $b_0C$. Explicitly, with the same assumptions as Cohen and Kaplan,
the asymptotic equations we need to solve are:
$$
\eqalignno{
[C'e^{2A}]' &= - {2\epsilon e^{2A} \over C} & (4.6a) \cr
[A'Ce^{2A}]' &= b_0C & (4.6b) \cr
{A'}^2 + {2A'C'\over C} &= b_0e^{-2A} - {\epsilon \over C^2} & (4.6c) \cr
}
$$

In what follows we will take $\epsilon \ll 1$, and also recall (3.19),
$b_0 <O(\epsilon^2)$.
We choose to write this analysis in a slightly different form. Let
$\rho = \int_0^r e^{-A} dr$, and denote ${d\over d\rho}$ by a dot.
Then, letting $f = {\dot A} + {\dot C}/C$, and $g={\dot C}/C$, we have
$$
\eqalignno{
f^2 &= b_0 + g^2 - {\epsilon e^{2A} \over C^2} & (4.7a) \cr
{\dot g} &= - fg - {2\epsilon e^{2A} \over C^2} 
= 2(f^2-g^2-b_0) -fg & (4.7b) \cr
}
$$
from (4.6a) and (4.6c). Now, differentiating (4.7b) and using (4.7a), we 
obtain
$$
{\dot f} = f^2 - b_0 - 2g^2
\eqno (4.7c)
$$
Thus we have reduced our (constrained) coupled second 
order differential equations
to a two-dimensional dynamical system, (4.7c) and (4.7b). Whether or
not (3.14) is admissible as an asymptotic solution for the global string
will now reduce to a question of whether or not the dynamical system
will asymptote the solution appropriate to (3.14) in phase space. 

First of all, consider $b_0=0$. Then there is just one fixed point,
$f=g=0$, which corresponds to $C$ and $A$ being constant, $C$ infinite.
Moreover, since
$$
\eqalign{
{\dot f} + {\dot g} &= (3f-4g) (f+g) \cr
{\dot f} - {\dot g} &= -f(f-g) \cr
}
\eqno (4.8)
$$
$f\pm g$ are separatrices in the phase plane, and the Cohen-Kaplan family
of solutions lie entirely in the upper quadrant. We can therefore see that
all CK solutions with $g>f>0$ initially asymptote $g\simeq -f \to \infty$,
i.e.\ ${\dot C}/C \simeq -{\dot A}/2 \to\infty$, which indeed agrees with
the CK solution.
A full phase diagram is shown in figure 2.
\midinsert \hskip 4truecm \epsfxsize=8truecm 
\epsfbox{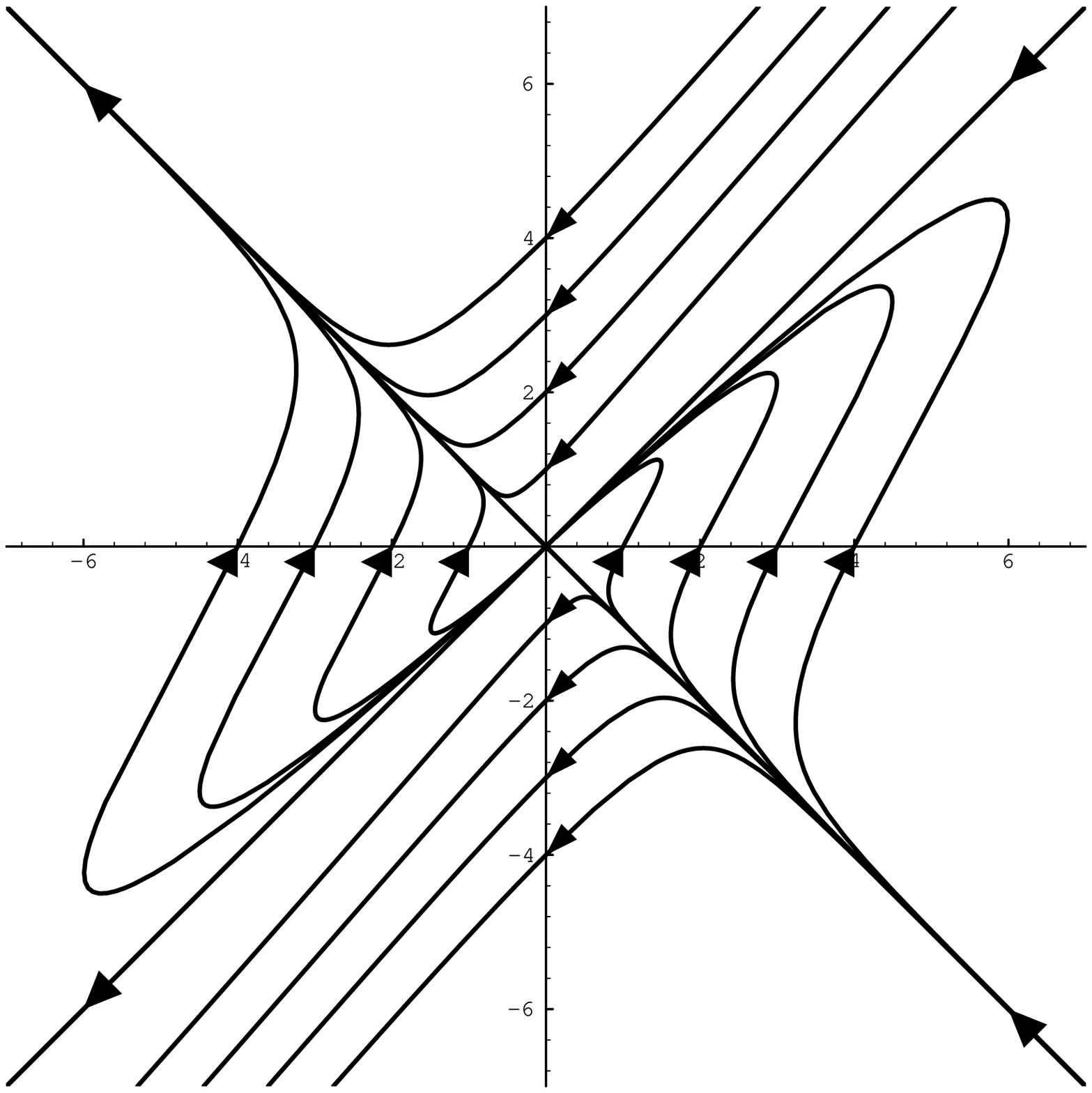} \medskip \hskip 2truecm
\vbox{ \hsize=11.5 truecm 
\noindent {\cat FIGURE (2): The phase plane of the static (CK) 
global string. }}
\endinsert

Now consider $b_0> 0$. We will first consider the general behaviour of
the dynamical system before investigating whether a solution with the 
asymptotic behaviour (3.14) is possible. First we rescale variables by
setting $t = \sqrt{b_0}\rho$, and $(f,g) = \sqrt{b_0}(x,y)$ to obtain
$$
\eqalignno{
{dx\over dt} &= x^2 - 2y^2 -1 & (4.9a) \cr
{dy\over dt} &= 2x^2 -2y^2 -2 -xy & (4.9b) \cr
}
$$
This system has two saddle points at $(\pm1,0)$ and two foci at
$(\pm\sqrt{2}, \pm 1/\sqrt{2})$. It is also straightforward to see 
that the hyperbolae
$x^2 - y^2 =1$ form separatrices in the plane.
The analysis is somewhat more complicated than the CK case, but a 
phase plane diagram for the system is shown in figure 3.
\midinsert \hskip 4truecm \epsfxsize=8truecm
\epsfbox{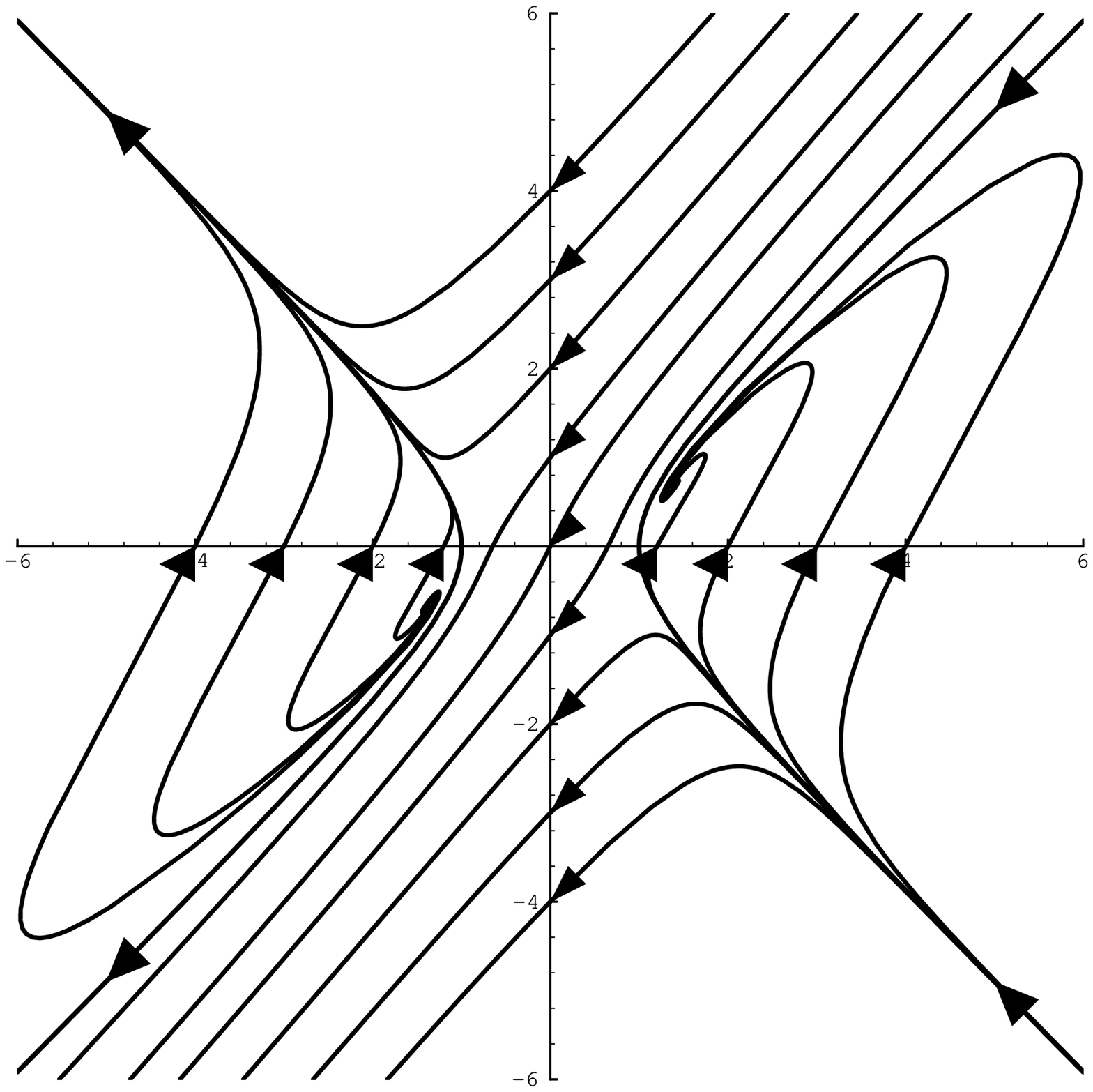} \medskip \hskip 2truecm
\vbox{ \hsize=11.5 truecm 
\noindent {\cat FIGURE (3): The phase plane of the time dependent
global string. }}
\endinsert

Now consider the critical point $(-1,0)$, this corresponds to $f=-\sqrt{b_0}$,
$g=0$. Inspecting  (4.7), we see that this corresponds to ${\dot C} =0$, 
${\dot A} = -\sqrt{b_0}$, $e^{2A}=0$, i.e.\ the asymptotic form of
(3.14). Therefore, asking whether a non-singular solution exists for the
global string reduces to asking whether a suitable
trajectory exists  which terminates
on the critical point $(-1,0)$ in the $(x,y)$-plane. 
Now, since $(-1,0)$ is a saddle point there does indeed exist a
unique trajectory approaching $(-1,0)$ -- the stable manifold, however,
the question is whether this trajectory is ``suitable'', i.e.
does it match on to the core of the global string? 
Therefore, we must now examine 
the initial conditions for the dynamical system, obtained from integrating 
out the full equations of motion, in order to see whether we can indeed fit
these initial conditions onto the required trajectory.

Remembering that $b_0<O(\epsilon^2)$, and letting $\rho_c =$O(1)
be a suitable value of $\rho$ representing the transition from core to
vacuum or the edge of the vortex, then
$$
{\dot A} (\rho_c) = ({\dot f} - {\dot g} )|_{\rho_c} \simeq
-{K\epsilon \over \rho_c}
\eqno (4.10)
$$
from (3.3a), where $K$ is as defined in (4.3). Then (4.7a) gives
$$
{g_0\over f_0} = {y_0\over x_0} = 1 + 2K^2\epsilon
\eqno (4.11)
$$
independent of $b_0$, and
$$
y_0 = {1\over 2K\rho_c\sqrt{b_0}}
\eqno (4.12)
$$
Clearly therefore, the trajectory approaching $(-1,0)$ in the $(x,y)$
plane will correspond to a global string if it intersects the line
$y=(1+2K^2\epsilon )x$ for some (large) $x>0$.

Now let us examine (4.9) for $x>0$. By observation, $y\in[1,2]$
at $x=0$ for the non-singular trajectory, 
therefore we can roughly bound $y$ by $[x+1, 2(x+1)]$
for general $x>0$. Therefore, as $t\to-\infty$ $x,y\to\infty$, thus
(4.9) asymptotes the CK system as $t\to-\infty$. The solution to leading
order for $x$ and $y$ can be written as
$$
y \simeq x \left [ 1 + {1\over 4\ln x} \right ]
\eqno (4.13)
$$
(a better approximation can be derived, but this will suffice for our 
purposes). In particular, note that for non-zero $\epsilon$, there
exists an $x_\epsilon$ such that $ 1 + 1/4\ln x < 1 + 2 K^2 \epsilon$
for all $x > x_\epsilon$, and hence the trajectory will indeed intersect 
$y = (1 + 2K^2 \epsilon)x$ at some value of $x$. This value of $x$ will then
determine $b_0$. An illustration of this process is shown in figure 4, where
we have inflated the value of $K^2\epsilon$ for the purposes of clarity.
For $8K^2\epsilon=1$, we obtain $b_0\simeq 10^{-3}$.
\midinsert \hskip 4truecm \epsfxsize=8truecm
\epsfbox{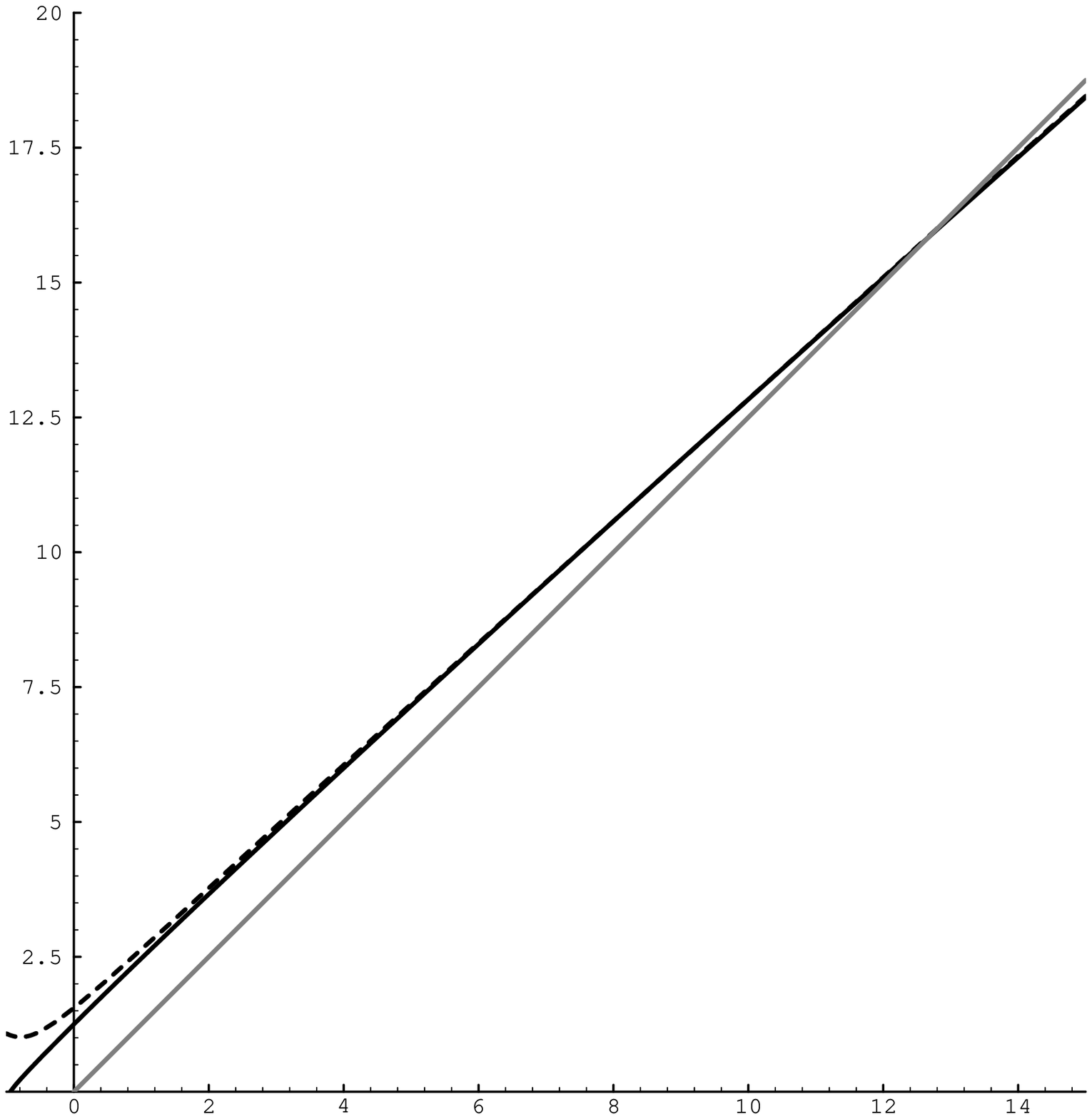} \medskip \hskip 2truecm
\vbox{ \hsize=11.5 truecm 
\noindent {\cat FIGURE (4): An illustration of the determination of $b_0$
from the intersection of the non-singular trajectory (shown as a 
continuous line) with $y=(1+2K^2\epsilon)x$ (shown as a grey line)
for the (rather large) value $2K^2\epsilon = 0.25$. The CK trajectory with
the same initial conditions is shown as a dashed line. The value of
$b_0$ for this solution is $b_0 = 1/(1028 K^2 \rho_c^2)\simeq 10^{-3}$.}}
\endinsert

Thus, what we have shown in this section is that by reducing the far-field
equations to a two-dimensional dynamical system, we are able 
to demonstrate the existence of a trajectory interpolating between 
the initial conditions at the edge of the vortex and the asymptotic solution
(3.14), the event horizon.

\vskip 5mm

\noindent{\bf 5. Discussion.}

\vskip 2mm

We now wish to explore the qualitative features of the solution represented
by the trajectory of section four. We will assume $\epsilon \ll 1$.
Clearly in the initial stages of the trajectory, for $x,y \gg 1$, we expect
a Cohen-Kaplan (CK) trajectory to be a good approximation, and only
when $x,y \simeq O(1)$ will the solution significantly differ from CK.
By dividing (4.9b) by (4.9a) we get the relation
$$
{dy\over dx} = 2 - {y(2y-x) \over 2y^2 +1 -x^2} >
2 - {y(2y-x) \over 2y^2 - x^2} \;\;\;{\rm for}\;\;\; y>x>0
\eqno (5.1)
$$
therefore, at all points in the upper half of the positive $(x,y)$ 
quadrant, the trajectories of the time dependent global string are
steeper than those of the static, CK, string (see figure 4). 
This allows us to put
an upper bound on $b_0$, since the CK trajectory starting at the
same initial conditions exterior to the core will always lie above the
real trajectory. Working in the $(f,g)$-plane, (4.4) gives
$$
\eqalign{
f &= {e^A\over C} \left [ -{1\over 4u} + {u \over u_0} \right ] \cr
g &= {e^A\over C} \left [ {1\over 4u} + {u \over u_0} \right ] \cr
}
\eqno (5.2)
$$
in terms of the variable $u$, and hence $f=0$ when $u = \sqrt{u_0}/2$.
At this point
$$
g \simeq \epsilon^{7\over8} e^{-{1\over2\epsilon}}
\eqno (5.3)
$$
Since $g = \sqrt{b_0}y$, and $y>1$ for this
CK trajectory, we see that
$$
b_0 < \epsilon^{7\over4} e^{-1/\epsilon}
\eqno (5.4)
$$
and indeed, since we expect the CK trajectory to be a good approximation
to the real trajectory until $(x,y) \simeq 1$ we expect $b_0$ 
not to be significantly less than this order of magnitude.
Therefore, the rate of expansion along the string, determined by $b_0$,
is minutely small. It is also interesting to note the proper radius
at which this transition from a CK solution to the asymptotic form (3.14)
occurs
$$
r_k = \int_{1/2\sqrt{u_0}}^{u_0} Cdu \simeq r_0 ( 1 - O(\epsilon^{{\half}}))
\eqno (5.5)
$$
For $r_k<r<r_0$ we expect that (3.14) will be a good approximation
to the spacetime.

From a cosmological point of view it is instructive to estimate these 
critical radii, $r_k$ and $r_0$, for a typical value of $\epsilon$
appropriate to a GUT string, $\epsilon = 10^{-6}$. This gives 
$r_0 = O(10^{100,000})$ with $r_k$ being of the same order. Even
allowing for the fact that $r_0$ is measured in units of string width, this
value is many many times the current Hubble radius, which is about $10^{52}$
in these units! Therefore, cosmologically speaking, not only is the
effect of the expansion negligible, but the gravitational field of
the string is not appreciably different from that of the singular CK
metric. In other words, our solution justifies the use of the CK
metric as an approximation to the gravity of a global string on 
intermediate scales.

It is perhaps more interesting to ask what happens if $\epsilon \simeq 1$.
Such heavy vortices have relevance to the topological inflation 
of Linde and Vilenkin [24,25]. Central to their argument is the
non-existence of static non-singular supermassive defect solutions,
otherwise supermassive defects would not inflate. Certainly, our solution
is non-static, however, it is not sufficiently non-static! The existence
of a non-singular metric of the general form (2.12) would mean that
topological inflation was not possible with global strings, since
the ``inflation'' in (2.12) is occurring only along the length of the
string.

Unfortunately, we cannot apply our results directly to this interesting
scenario, since for $\epsilon\simeq1$ the analysis of the previous
section cannot be straightforwardly applied. It will still be true
that there will exist a trajectory with the correct 
asymptotic behaviour of (3.14), but the appropriate initial conditions
can no longer be approximated since we can no longer use the 
flat space solutions to estimate ${\dot A}(\rho_c)$. Indeed,
whether it is appropriate to be performing an asymptotic ($X\simeq1$)
analysis in such a strongly coupled r\'egime is also questionable.
It is probably necessary to perform a more detailed numerical
investigation in order to resolve this issue.

Finally, it is interesting to return to the question of detecting axion hair.
This hair will only be detectable if there exist Euclidean vortices on the 
event horizons of Schwarzschild black holes which still
have asymptotically flat geometries. There are two main reasons why we do
not expect the solutions presented here to satisfy this criterion. First,
although the solution is non-singular, it does have an event horizon,
i.e.\ a strong asymptotic effect. This tends to indicate that
Euclidean global vortices will not be asymptotically flat. However,
the second objection is more troubling. In deriving the metric (2.12),
and the ensuing analysis, implicit use was made of the non-compactness
of the worldsheet directions $z$ and $t$. In particular, the existence
of a continuum of choices for what is essentially an eigenvalue, $b_0$,
was crucial to our argument. For the global vortex on a black hole,
the worldsheet $z$ and $t$ are replaced by $\theta $ and $\phi$, which 
coordinatise a compact manifold. This means that not only would the
corresponding metric have angular dependence, but in addition, we might
expect a discrete spectrum of eigenvalues in an analogous
metric to (2.12), which means that we can no longer continuously vary
$b_0$ to hit the right non-singular trajectory. In other words
Euclidean global vortices are quite probably singular no
matter what one does. Axion hair, for the moment, must
remain undetectable.

\vskip 5mm

\noindent{\bf Acknowledgements.}

\vskip 2mm

It is a pleasure to thank Caroline Santos for useful discussions on
dynamical systems. This work was supported by a Royal Society University
Research Fellowship.

\vskip 5mm

\noindent{ \bf References.}

\vskip 2mm

\nref
R.H.Brandenberger, {\it Modern Cosmology and Structure Formation}
astro-ph/9411049.
 
\nref
M.B.Hindmarsh and T.W.B.Kibble, \rprog 58 477 1995. [hep-ph/9411342]
 
\nref
A.Vilenkin and E.P.S.Shellard, {\it Cosmic strings and other 
Topological Defects} (Cambridge Univ. Press, Cambridge, 1994).
 
\nref
A.Vilenkin, \prd 23 852 1981.
 
\nref
D.Garfinkle, \prd 32 1323 1985.
 
\nref
R.Gregory, \prl 59 740 1987.
 
\nref
M.Ortiz, \prd 45 2586 1992.

\nref
K.Lee, V.P.Nair, and E.J.Weinberg, \prd 45 2751 1992. [hep-th/9112008]
 
\nref
A.G.Cohen and D.B.Kaplan, \plb 215 67 1988.

\nref
M.Barriola and A.Vilenkin, \prl 63 341 1989.

\nref
R.Gregory, \plb 215 663 1988.
 
\nref
G.Gibbons, M.Ortiz and F.Ruiz, \prd 39 1546 1989.
 
\nref
A.Vilenkin, \plb 133 177 1983.

\nref
J.Ipser and P.Sikivie, \prd 30 712 1984.

\nref
M.Bowick, S.Giddings, J.Harvey, G.Horowitz and A.Strominger, \prl 61 2823 1988.
 
\nref
L.Krauss and F.Wilczek, \prl 62 1221 1989.
 
\nref
S.Coleman, J.Preskill and F.Wilczek, \mpla 6 1631 1991.
\prl 67 1975 1991. \npb 378 175 1992. [hep-th/9201059]

\nref
H.F.Dowker, R.Gregory and J.Traschen, \prd 45 2762 1992. [hep-th/9112065]

\nref
J.A.Stein-Schabes and A.B.Burd, \prd 37 1401 1988.
 
\nref
R.Gregory, \prd 39 2108 1989.
 
\nref
E.Shaver and K.Lake, \prd 40 3287 1989.

\nref
A.Banerjee, N.Banerjee and A.A.Sen, \prd 53 5508 1996.

\nref
D.Harari and A.Polychronakos, \plb 240 55 1990.

\nref 
A.D.Linde, \plb 327 208 1994. [astro-ph/9402031]

\nref
A.Vilenkin, \prl 72 3137 1994. [hep-th/9402085]

\bye
\vfill\eject

\noindent{\bf Figure Captions}
\vskip 3mm
 
FIGURE (1): The ``Kruskal'' diagram of the global
string spacetime in $\{X,Y,T\}$ coordinates. Region II, exterior to 
the cone, corresponds to the global string spacetime interior to the
event horizon. Regions I and I' correspond to the spacetime exterior
to the event horizon reached by future and past pointing null geodesics
respectively.
 
\vskip 2mm
 
FIGURE (2): The phase plane of the static (CK) 
global string.
 
\vskip 2mm
 
FIGURE (3): The phase plane of the time dependent
global string.
 
\vskip 2mm
 
FIGURE (4): An illustration of the determination of $b_0$
from the intersection of the non-singular trajectory (shown as a 
continuous line) with $y=(1+2K^2\epsilon)x$ (shown as a grey line)
for the (rather large) value $2K^2\epsilon = 0.25$. The CK trajectory with
the same initial conditions is shown as a dashed line. The value of
$b_0$ for this solution is $b_0 = 1/(1028 K^2 \rho_c^2)\simeq 10^{-3}$.

\bye